\documentclass[prl,aps,epsf,twocolumn,nofootinbib,notitlepage,showpacs,10pt]{revtex4-1}
\usepackage{bm}% bold maths
\usepackage{amsmath, verbatim,graphicx}
\usepackage{amsfonts}
\newcommand{\ket}[1]{| {#1} \rangle}

\begin{document}

\title{A tree tensor network approach to simulating Shor's algorithm}
\author{Eugene Dumitrescu$^{1,2}$}
\affiliation{$^1$Quantum Computing Institute, Oak Ridge National Laboratory, Oak Ridge, TN 37831 \\
$^2$ Bredesen Center for Interdisciplinary Research, University of Tennessee, Knoxville, TN 37996}

\begin{abstract}
Simulating quantum systems constructively furthers our understanding of qualitative and quantitative features which may be analytically intractable.
In this letter, we directly simulate and explore the entanglement structure present in a paradigmatic example of quantum information: Shor's wavefunction. 
The methodology employed is a dynamical tensor network which is initially constructed as a tree tensor network, inspired by the modular exponentiation quantum circuit, and later efficiently mapped to a matrix product state.
Utilizing the Schmidt number as a local entanglement metric, our construction explicitly captures the wavefunction's non-local entanglement structure and an entanglement scaling relation is discovered.
Specifically, we see that entanglement across a bipartition grows exponentially in the number of qubits before saturating at a critical scale which is proportional to the modular periodicity. 
\end{abstract}

\maketitle

%% Required disclaimer
\footnotetext{This manuscript has been authored by UT-Battelle, LLC, under Contract No. DE-AC0500OR22725 with the U.S. Department of Energy.
The United States Government retains and the publisher, by accepting the article for publication, acknowledges that the United States Government retains a non-exclusive, paid-up, irrevocable, world-wide license to publish or reproduce the published form of this manuscript, or allow others to do so, for the United States Government purposes. The Department of Energy will provide public access to these results of federally sponsored research in accordance with the DOE Public Access Plan.}

%introduction
{\indent{\em Introduction.}}---
Tensor networks are graphical data structures consisting of nodal tensors with indexed edges which represent physical and virtual degrees of freedom. 
The graphical connectivity and dimensionality of the virtual degrees of freedom encode entanglement in a localized manner. 
Arising naturally from the need to efficiently decompose entangled many-body wavefunctions, tensor networks have been tremendously successful in numerically finding ground states of local Hamiltonians in low-dimensional condensed matter systems \cite{White_DMRG, Orus_review} and also have applications in simulating some quantum circuits \cite{Shi_2006, Markov_Shi, Biamonte_2013, Wang_2015}.

Along with the ability to calculate local observables, tensor network representations of many-body wavefunctions provide explicit insight into entanglement structure, which differs vastly based on system dimensionality and criticality\cite{Hastings_2011, Schollwoeck_2011, Pirvu_2012}. 
Aside from finding ground states, tensor networks have recently successfully simulated the evolution of open quantum systems\cite{Werner_2014, Schroder_2016}.
Despite these benefits, finding an appropriate and efficient tensor network representation for a physical system is not a simple task. 
This is especially true for higher dimensional systems, for which no general efficient methods are known \cite{Orus_review}. 

In contrast to condensed matter states residing on a lattice, dimensionality or local geometry are ill-defined quantities for quantum information theoretic states which are generated by logical quantum circuits.
Motivated by the algorithmic structure and permutational invariance of qubits involved in the modular exponentiation step, we develop a bipartite tree tensor network (TTN) structure which naturally simulates states generated through Shor's algorithm.
Inspecting the entanglement via the TTN decomposition demonstrates a clear and manifest area law violation which we discussed in detail. 
Additionally, our TTN state is efficiently converted into a matrix product state (MPS) tensor network so that the output state can be subjected to the quantum Fourier transform component of the algorithm. 
The tensor networks we develop, although not fully efficient, are useful for representing moderate size systems, with up to 39 qubit wavefunctions constructed on a laptop computer. 
Our dynamical tree tensor network methodology is generally applicable to many familiar information theoretic bipartite quantum states, e.g. those generated in dual register systems by the quantum phase estimation and hidden subgroup algorithms\cite{Nielsen}. 

\begin{figure}[htb!]
\begin{center}
\includegraphics[width = \columnwidth]{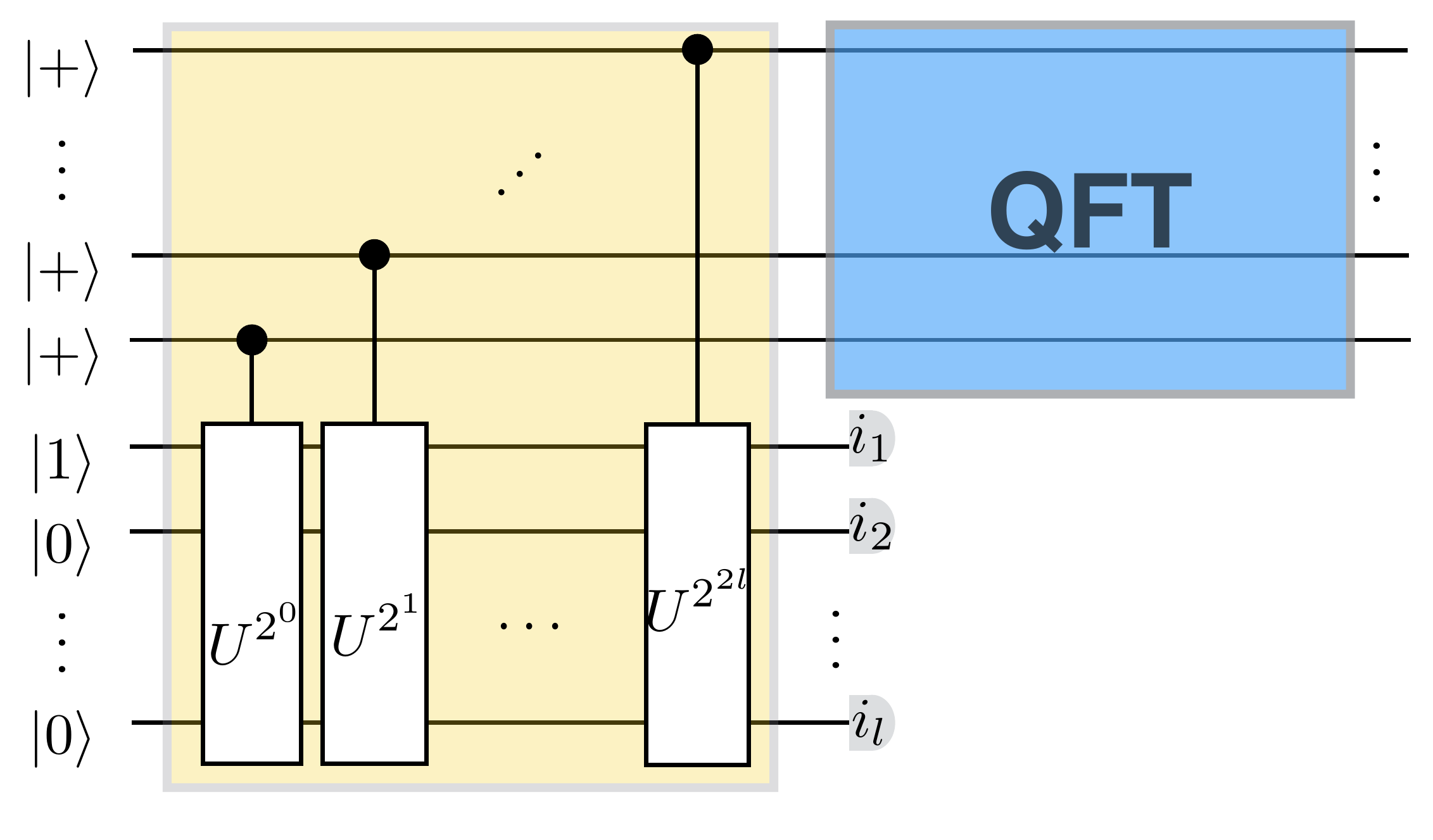}
\caption{Schematic of Shor's algorithm with the ME (QFT) components highlighted in yellow (blue) boxes. 
Eq.~\ref{eq:order_schmidt} describes the state of the system one time-step before the bottom register measurements proceeding the ME sub-circuit.}
\label{fig:Shor}
\end{center}
\end{figure}

% Shors algorithm and wavefunction
{\indent{\em Shor's Wavefunction.}}---
We first outline the logical operations comprising Shor's algorithm \cite{Shor} in order to develop an intuition for the form of an appropriate tensor network. 
To factor a natural number $N = pq$, with $p,q$ large prime numbers, we draw a random integer $x \in \mathbb{Z}_N$ and use Shor's algorithm \cite{Shor} to find the characteristic modular periodicity $r$ given by $x^r \mod N = 1$.
Assuming that $gcd(x,N) = 1$ -- in the unlikely case that $x=p$ or $q$, $N$ is trivially factored -- one initializes $2l$ ($l$) qubit top (bottom) register, for a total of $3 l$ qubits, where $l = \log_2(N)$ is the number of bits required to represent $N$.
The top register is initialized into the product state $|+\rangle^{\otimes 2l} = 1/(\sqrt{2})^{2l}\sum_{i=0}^{2^{2l}-1} |i\rangle$ while the bottom register is initialized as $\ket{1}$ which is the integer basis representation of the computational basis state $\ket{0,0,...,0,1}$. 
We represent the {\em bottom} register in the integer basis for the remainder of this letter.
The composite initial product state is thus $|\Psi_i \rangle = |+\rangle^{\otimes 2l}_{top} \otimes |1\rangle_{bot}$. 

The modular exponentiation (ME) unit (see Fig.~\ref{fig:Shor} yellow box) entangles each top register qubit with the entire bottom register via controlled modular multiplication operators $U \equiv U(x,N)$. 
The operator $U$ is a rank $2$ tensor with dimensions $2^l \times 2^l$ and matrix elements satisfying $U \ket{b} = \ket{x b \mod{N}}$. 
Powers of $U^{2^i}$ are generated by $i$ iterative matrix multiplications. 
Upon application of the last controlled operator, the state reads $\ket{\Psi} = 2^{-l} \sum_{i = 0}^{2^{2l}-1} | i \rangle \otimes |x^i \mod N\rangle$. 
Because $x^r \mod N = 1$ we may group together like bottom register basis vectors and write the state as 
\begin{equation}
\label{eq:order_schmidt}
\ket{\Psi}= \frac{1}{\sqrt{r}} \sum_{i=0}^{r-1}  \left( \sum_{j=0}^{\lceil(2^{2l}-1)/r \rceil} |jr+i \rangle \right) \otimes | x^i \mod N \rangle. 
\end{equation}

% state complexity
{\indent{\em Bipartite tensor network complexity.}}---
Eq.~\ref{eq:order_schmidt} is by definition a bipartite Schmidt decomposition between the two registers and reveals several interesting features. 
We see $|\Psi\rangle$ is $r$-entangled across the bipartition, that is, the Schmidt coefficient $1/\sqrt{r}$ appears $r$ times. 
In the worst case $r \sim \mathcal{O}(N)$ so the inter-register entanglement scales exponentially in the number of qubits $l$ \cite{Orus_2004}.
The equality of all Schmidt coefficients also foreshadows different correlations scaling compared to ground states of local Hamiltonians, which have exponentially (or power-law) decaying correlations.
Eq.~\ref{eq:order_schmidt} also demonstrates that it is natural to decompose $|\Psi\rangle$ across the inter-register bipartition, and we shall retain this feature in our tree network.

The quantum Fourier transformation is known to be efficiently simulable \cite{Aharonov_2006, Yoran_2007}, suggesting that the non-trivial part of the computation occurs during the modular exponentiation step. 
We thus pose the following question: what are the entanglement properties of the basis state $\sum_{j=0}^{\lceil(2^{2l}-1)/r \rceil} |jr+i \rangle$? 
The top register qubits are clearly entangled with one another via their interaction with the bottom register. 
We therefore know from Eq.~\ref{eq:order_schmidt} that $r$ sets an upper bound on the amount of entanglement. 
Below we elucidate the entanglement properties of the top register basis states by developing a tensor network representation whose geometry is consistent with the inter-register bipartition and, more importantly, by the permutational invariance of the top register qubits. 

A first attempt at a tensor network was performed in Ref.~\citenum{Wang_2015}, which treats the bottom register as a qudit lying at one end of a MPS. 
The ME algorithm was performed by contracting two-local controlled modular multiplication gates along with a series of swap gates. 
In doing so, the complexity of storing the state is reduced from $\mathcal{O}(2^{3l}) \rightarrow \mathcal{O}(2^{l} r) + const$ with the a constant given by $ \sum_j 2 * d^{(j)}_{l} d^{(j)}_{r}$ where $d^{(j)}_{l(r)}$ are the virtual bond dimensions to the right and left of the $j$th top register qubit. 
While this approach was successful in simulating Shor's algorithm, artificially large virtual bond dimensions were generated by successive swap operations. 
This leads to a situation where $d^{(j)}_{l(r)} = r$ for many bonds when, as we shall see, few virtual bonds of that size are necessary. 

\begin{figure}[htb!]
\begin{center}
\includegraphics[width = \columnwidth]{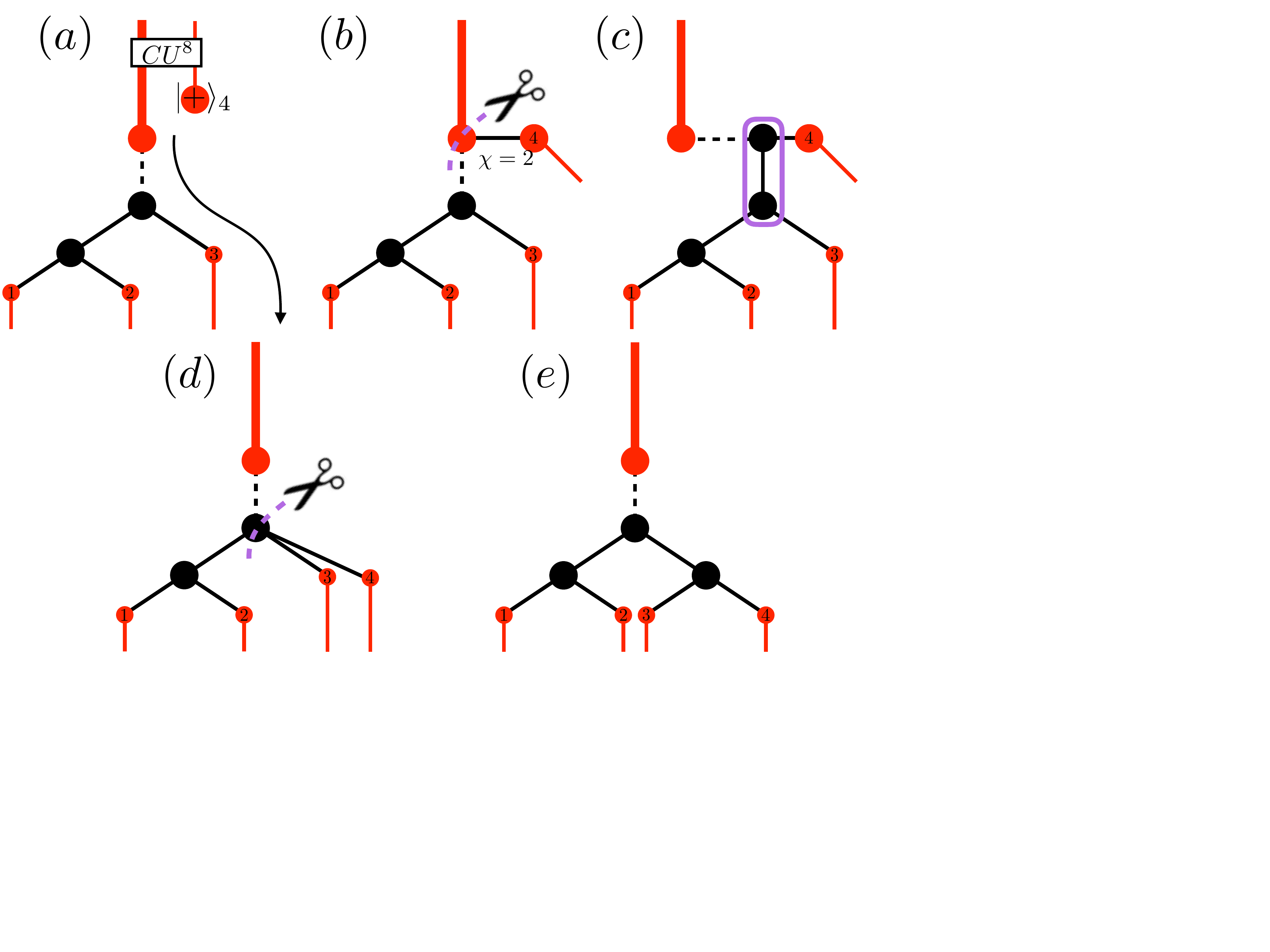}
\caption{Intermediate virtual networks for (b-d) between the application of a controlled modular multiplication gate (a) and the updated tree network (e).
Red (black) lines denote physical (virtual) degrees of freedom.
The thick red line at the base of the tree denotes the $2^{l}$ dimensional qudit bottom register state. 
Top register qubits are represented by thin red lines at the bottom of the tree. 
The dashed black line denotes the bipartition defined by Eq.~\ref{eq:order_schmidt}. 
The scissors icon and dashed purple lines (purple box) denote the bipartition chosen for tensor decompositions (contraction) generating the next tree configuration.} 
\label{fig:Tree_Update}
\end{center}
\end{figure}

{\indent{\em Tree Generation Algorithm.}}--- 
We now introduce a natural tensor network which maintains the inter-register bipartition and distributes entanglement in an unbiased manner.
This network is dynamically constructed by following the ME sub-circuit, as shown in Fig.~\ref{fig:Shor}, with intermediate virtual updates performed, as shown in Fig.~\ref{fig:Tree_Update} and discussed below, in between circuit operations. 
Our construction algorithm goes as follows.
(i) Contract the $i$th controlled $U^{2^i}$ operator with the $i$th single qubit $\ket{+}$ tensor and the current bottom qudit state as shown in panel $(a)$.  
(ii) Perform internal operations updating and generating virtual indices. 
This cascades the $i$th qubit from the tree root (i.e. directly connected to the bottom register) to a new bottom tree branch as illustrated in panels $(b-e)$. 
Repeat the procedure for all qubits indexed by $i \in  \mathbb{Z}_{2l}$.

\begin{figure}[htb!]
\begin{center}
\includegraphics[width = \columnwidth]{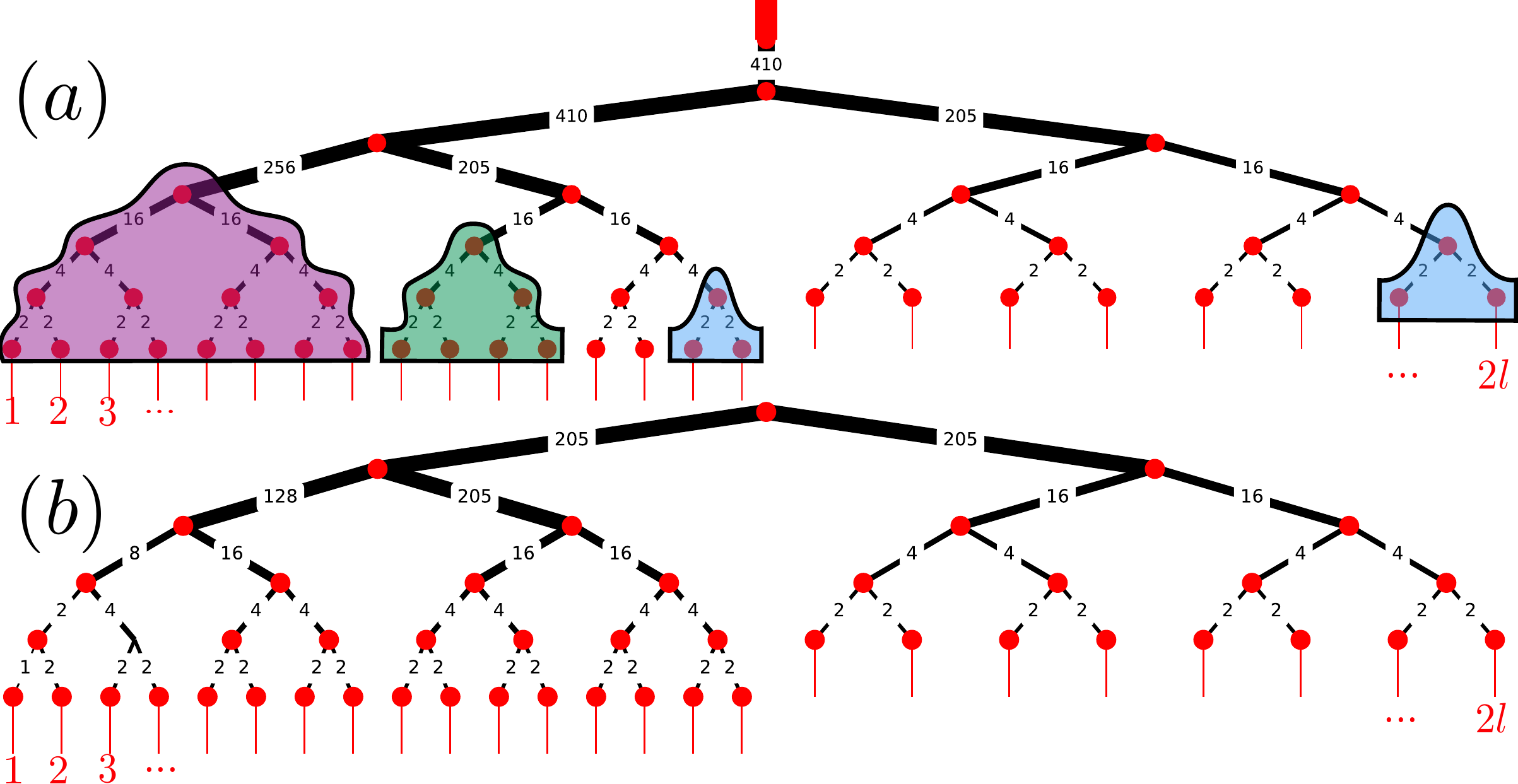}
\caption{Tree tensor network decomposition of $\ket{\Psi}$ from Eq.~\ref{eq:order_schmidt} for an $l=12$ size system with $N = 3403= 41\times 83$, $x=346$, and cyclic order $r = 410$.
Edges widths are $\log(d_i)+1$ where $d_i$ is the $i$th bond dimension which has been labeled.
Top qubit (qudit) physical degrees of freedom appear as red edges along the bottom (top). 
Highlighted regions are entangled to their complement by a common parent branch whose dimension follows Eq.~\ref{eq:S}}
\label{fig:Tree}
\end{center}
\end{figure}

The internal updates consists of the following steps:
(i) After applying a controlled $U^{2^i}$ gate, a SVD separates the $i$th qubit from the root qudit (Fig.~\ref{fig:Tree_Update} panel $(b)$). 
The $i$th qubit is maximally entangled with the bottom register via a $\chi = 2$ dimensional auxiliary edge with singular values $(\frac{1}{\sqrt{2}},\frac{1}{\sqrt{2}})$
(ii) Generate the new inter-register entanglement bond by performing an SVD between the bottom register and its local complement formed by the union of the new qubit and the previous tree root as indicated by the dashed purple line in panel $(b)$. 
Recall that this bond's dimensionality eventually saturates at $r$.
(iii) The tensors encircled by the purple box in panel $(c)$ are contracted in order to `lower' the qubit through the tree before, 
(iv) another SVD along a bipartition, which is chosen to direct the qubit through a specific path, is performed. 
Step (iv) is identical to step (ii) but occurs further down the tree. 
Repeat steps (iii,iv) until each qubit settles into its final location at the bottom of the tree.
An example of a final tree configuration is illustrated in Fig.~\ref{fig:Tree}. 

Note that the choice of a binary tree is arbitrary and that the entanglement features discussed in the next section hold for aribtrary tree data structure. 
Also note that the number of virtual updates cascading the $i$th qubit is clearly upper bounded by final tree depth. 
Since the final tree depth is logarithmic in the number of qubits, that is with depth $\lceil \log(2 l) \rceil$, the tree construction procedure is efficient. 
This trade-off can be compared to that in an MPS based simulation for which at least $2l$ swaps are performed. 
Thus, a logarithmic number of updates to generate an unbiased and natural representation of the state is well justified and we now discuss the emergent entanglement properties.  

{\indent{\em Entanglement Features.}}---
After cascading all qubits, the tree tensor network exactly encodes the wavefunction appearing in Eq.~\ref{eq:order_schmidt}. 
The entanglement structure for the top register, not apparent in Eq.~\ref{eq:order_schmidt}, is now revealed by the holographic dimension \cite{Orus_2014} constructed by our virtual updates. 
In our analysis we use the Schmidt number, given by the bond dimension of the virtual index connecting bipartitions, as the metric for shared entanglement. 
This is an appropriate metric because, unlike ground states of local Hamiltonians which have exponential or power-law decaying Schmidt coefficients, the Schmidt coefficients are {\em equal} in magnitude.
Thus the Schmidt number completely describes the entanglement which can therefore be visualized as done in Fig.~\ref{fig:Tree}, where the drawn bonds are weighted as $\log_2{(d_i)}+1$, where $d_i$ is the local bond dimension which is also labeled. 

At the bottom tree level where qubits first connect to their parent branches all qubits are {\em maximally} entangled to the rest of the network, with equal Schmidt coefficients $\lambda_0 = \lambda_1 = \frac{1}{\sqrt{2}}$. 
At the next level, all pairs of qubits (e.g. Fig.~\ref{fig:Tree} blue highlighted qubits) are still maximally entangled with their complement, i.e. with degenerate Schmidt coefficients $\lambda_i = 0.5$ for $i = (0,1,2,3)$.
This trend, with clusters of $2^n$ qubits maximally entangled to the rest of the state by $2^n$ identical Schmidt coefficients (e.g. Fig.~\ref{fig:Tree} cluster of 4 green highlighted qubits, and so on) continues up to a critical size, at which point the entanglement rapidly saturates. 

The qubit cluster size at which the entanglement scaling saturates depends on which qubits are selected and is either $l_r = \lceil \log_2(r) \rceil$ or $l_{\tilde{r}} = \lceil \log_2(\tilde{r}) \rceil$, where $\tilde{r} = r/2^m$ and $m$ is the largest integer such that $2^m$ divides $r$.
The critical length scale, specific to the choice of $N$ and $x$ determining $r$, can be understood by the following arguments. 
The $r$ dimensional tree root bond mediates the $r$-fold entanglement across the register bipartition as per Eq.~\ref{eq:order_schmidt}. 
Further, $r$ constrains the intra-register entanglement because all entanglement between qubits, stored in the bulk, was generated by controlled modular multiplication gates acting solely on the qudit. 
Descending from the tree root, bond dimensions decrease from $r$ to either $\tilde{r}$, or powers of 2 less than $r,\tilde{r}$. 
An example is provided by Fig.~\ref{fig:Tree} (a) where we have illustrated the final tree generated for $N = 3403$. 
Thus the entanglement scales as:
\begin{equation}
\label{eq:S}
S=
    \begin{cases}
      2^n, & \text{if}\ n<l_{r(\tilde{r})} \\
      r(\tilde{r}), & \text{otherwise}
    \end{cases}
\end{equation}
where $n$ refers to the number of qubits and the saturation dimension, $r$ vs $\tilde{r}$, depends on whether qubits belong to the first $l_r$ qubits (as seen from left to right in Fig.~\ref{fig:Tree}) or to the remainder of the register.
Eq.~\ref{eq:S} defines a class states whose entanglement is reminiscent of quantum error correcting codes with entanglement set by a distance $d$ for $[[n,k,d]]$ codes \cite{Gottesman_97}.

We now address the seemingly strange feature of why the first $l_r$ qubits are more entangled than the others.
Note modular exponentiation with at least $l_r$ qubits is needed to generate the $r$-fold basis vectors on either side of the register bipartition. 
Modular exponentiation with the remaining qubits extends the orthonormal basis vectors $|jr+i \rangle$ as the Hilbert space grows, leaving the bipartite entanglement at order $r$. 
To understand why the latter qubits are less entangled than the former, consider a single orthonormal Schmidt vector $\sum_{j=0}^{\lceil(2^{2l}-1)/r \rceil} |jr+i \rangle$. 
The qubits are projected into such a state upon the measurement of the bottom register qudit. 
Since $r = 2^m \tilde{r}$ (if r is odd the algorithm restarts with a different $x$) and $i<r$, the last $m$ bits for each $jr+i \rangle$ are the identical. 
Qubits $1-m$ are therefore disentangled from the remaining state, and the remaining entanglement now follows the scaling law in Eq.\ref{eq:S} saturating at $\tilde{r}$. 
Fig~\ref{fig:Tree} (b) illustrates this point by re-plotting the tree after a qudit measurement and bond updates are performed. Note the changes in the bond dimensions along the left side of the tree. 

\begin{figure}[htb!]
\begin{center}
\includegraphics[width = \columnwidth]{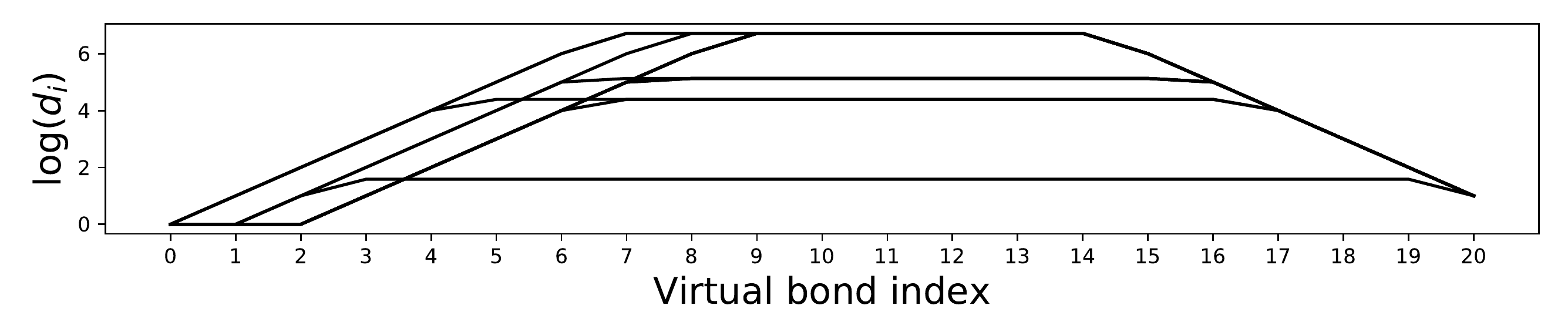}
\caption{Entanglement as quantified by MPS auxiliary bond dimensions for an $N=1763, l=12$ system .
The virtual bond dimensions and entanglement grows exponentially between adjacent bipartitions for the first few sites from both the left and right end points and is saturated in the bulk of the system by the characteristic modular periodicity $\tilde{r}$.}
\label{fig:MPS_Results}
\end{center}
\end{figure}

{\indent{\em MPS conversion and interpretation.}}---
We now briefly comment on the conversion of a tree network to a MPS network, which is useful in simulating Shor's algorithm in its entirety, due to an efficient representation of the QFT component for MPS systems \cite{Aharonov_2006, Yoran_2007, Wang_2015}. 
A computational basis state measurement $\mathcal{M}_i$ (illustrated by the nodes indexed $(i_1,...,i_l)$ in Fig.~\ref{fig:Shor}) projects the qudit register onto a basis state $\ket{i}$ from Eq.~\ref{eq:order_schmidt}. 
We can probabilistically simulate this measurement by contracting the bottom register tensor with basis states $\ket{i'}$  to find non-zero matrix elements and then choose one randomly, with each measurement outcome being equally probable.

A series of virtual updates now reduce the tree network into the MPS form. 
The first (leftmost) first qubit in Fig~\ref{fig:Tree} is already in the MPS form because a single tensor connects a physical and virtual degree of freedom. 
We proceed by selecting the next qubit and contracting its parent bonds until it connects to the virtual degree of freedom to the right of qubit 1. 
After contracting the parent bonds of qubit 3, a decomposition is performed to its right in order to generate a virtual MPS bond not connected to qubit 4. 
This procedure is iteratively repeated for all remaining qubits, at which point the resulting network is an MPS. 
Again, the number of updates is bounded by the number of qubits and the logarithmic tree depth. 

Inspecting the MPS virtual Schmidt coefficients provides a complimentary perspective to that provided by Fig.~\ref{fig:Tree}. 
In Fig.~\ref{fig:MPS_Results} we plot the virtual bond dimensions across the MPS network for 24 simulations involving the same $N=1763$ with $x$ randomly chosen. 
Many distinct $x$ balues share the same order $r_x$, so their entanglement spectrums are superimposed. 
The MPS spectrum in Fig.~\ref{fig:MPS_Results} verifies the entanglement scaling described by Eq.~\ref{eq:S}, namely, (i) the first $m$ qubits are disentangled, (ii) entanglement grows exponentially up to a critical length scale, and (ii) the entanglement saturates at the scale $\tilde{r} \equiv r/2^m$. 

{\indent{\em Discussion and conclusion.}}---
% some edits
In this letter we have explicitly constructed a bipartite tree tensor network naturally representing Shor's wavefunction and used the resulting decompositions to elucidate the multi-partite entanglement properties.
Our representation is generated in a circuit-like manner, i.e. by contracting the modular multiplication gates in series, along with a logarithmic number of decompositions between successive gates.  
The tensor network developed is fully compatible with the MPS formalism, allowing for an efficient translation between the two tensor network representations. 
Our method allows us to examine the entanglement scaling, as quantified by the Schmidt coefficients, for specific instances of Shor's algorithm.
One distinguishing feature is that the Schmidt coefficients are constant across all decompositions and irrespective of their number and magnitude.
The Schmidt rank across system bipartitions involving regions smaller than a critical threshold scale exponentially in the number of qubits enclosed, i.e. regions containing less than  $\log_2(\tilde{r})$, where $\tilde{r} \propto r$, qubits are maximally entangled to their compliment. 
In contrast, the entanglement for bipartitions involving larger regions saturates at $\tilde{r}$.

Our explicit entanglement analysis, for wavefunctions leading to quantum speedups, offers new insight into quantum complexity theory. 
We have also reiterated the difficulty in classically simulating Shor's algorithm by demonstrating that conventional tensor network approximations cannot be made due to the equality of Schmidt coefficients. 
This contrasts condensed matter simulations in which truncation is a crucial numerical approximation, for example, underlying the efficiency of the 1D DMRG algorithm. 
It is clear that similar truncation techniques cannot apply to our system, so it is interesting to consider alternative approximations. 
As the computational bottleneck was provided by the full representation of the bottom register qudit, we briefly consider approximations and generalizations of our TTN to reduce this cost and allow the simulation of larger systems.
A first approximation could be performed by simply eliminating a set ratio of the bottom register basis vectors. 
This will deform the modular multiplication operators and lead to a state similar to Eq.~\ref{eq:order_schmidt}, except coefficients having support on, and being generated via the eliminated basis vectors will vanish. 
The modular periodicity $r$ will be partially encoded in this state and may still be extracted after the QFT.

A  completely different approach would be to explicitly expand the bottom register qudit into its own tree such that the bipartite tree tensor network consists of two opposing trees.  
The structure of the bottom register tree would be of tremendous interest, and would likely provide insight into the complexity of the ME protocol.
For this to work, the application of a modular multiplication gate should be broken down into its primitive components, i.e. modular addition, etc. \cite{vanMeter_2005, Pavlidis_2013} and both trees updated dynamically. 
Assuming the entire algorithm is not classically efficiently stimulable, an additional computational bottleneck may then arise from the presumably large (possibly exponential) number of virtual updates propagating the entanglement  onto the inter-register bond.  

{\indent{\em Acknowledgements.}}---
E. D. would like to thank R. Bennink for careful reading of the manuscript. 
Research sponsored by the Intelligence Community Postdoctoral Research Fellowship and the Laboratory Directed Research and Development Program of Oak Ridge National Laboratory, managed by UT-Battelle, LLC, for the U.S. Department of Energy. This manuscript has been authored by UT-Battelle, LLC, under Contract No. DE-AC0500OR22725 with the U.S. Department of Energy.


\begin{thebibliography}{10}

% General TNs

\bibitem{White_DMRG} ``Density matrix formulation for quantum renormalization groups'', S. R. White, Phys. Rev. Lett. {\bf 69}, 2863 (1992)

\bibitem{Orus_review} ``A Practical Introduction to Tensor Networks: Matrix Product States and Projected Entangled Pair States'', R. Orus, Annals of Physics {\bf 349} 117-158, (2014)

\bibitem{Biamonte_2013} ``Solving search problems by strongly simulating quantum circuits'' T. H. Johnson, J. D. Biamonte, S. R. Clark and D. Jaksch, Scientific Reports {\bf 3}, Article number: 1235 (2013)


% QI TNs

\bibitem{Shi_2006} ``Classical simulation of quantum many-body systems with a tree tensor network'', Y.-Y. Shi, L.-M. Duan, and G. Vidal, Phys. Rev. A {\bf 74}, 022320 (2006)

\bibitem{Markov_Shi} ``Simulating Quantum Computation by Contracting Tensor Networks'', I. L. Markov and Y. Shi, SIAM Journal on Computing, 38(3):963-981 (2008)

\bibitem{Wang_2015} ``Simulations of Shor's Algorithm using Matrix Product States, D. S. Wang, C. D. Hill, L. C. L. Hollenberg'', arXiv:1501.07644v1 [quant-ph] (2015)

%critical systems

\bibitem{Hastings_2011} ``An Area Law for One Dimensional Quantum Systems'', M. B. Hastings, JSTAT, P08024 (2007)

\bibitem{Schollwoeck_2011} ``The density-matrix renormalization group in the age of matrix product states'', U. Schollwoeck, Annals of Physics 326, 96 (2011)

\bibitem{Pirvu_2012} ``Matrix product states for critical spin chains: Finite-size versus finite-entanglement scaling'', B. Pirvu, G. Vidal, F. Verstraete, and L. Tagliacozzo, Phys. Rev. B {\bf 86}, 075117 (2012)

% open quantum systems

\bibitem{Werner_2014} ``A positive tensor network approach for simulating open quantum many-body systems'', A. H. Werner, D. Jaschke, P. Silvi, M. Kliesch, T. Calarco, J. Eisert, S. Montangero, Phys. Rev. Lett. {\bf 116}, 237201 (2016)

\bibitem{Schroder_2016} ``Simulating open quantum dynamics with time-dependent variational matrix product states: Towards microscopic correlation of environment dynamics and reduced system evolution'', F. A. Y. N. Schröder and A. W. Chin, Phys. Rev. B 93, 075105 (2016)

 %QFT

\bibitem{Aharonov_2006} ``The quantum FFT can be classically simulated'', D. Aharonov, Z. Landau, J. Makowsky, arXiv:quant-ph/0611156 (2006)

\bibitem{Yoran_2007} ``Efficient classical simulation of the approximate quantum Fourier transform'', N. Yoran and A. J. Short, Phys. Rev. A {\bf76}, 042321 (2007)


%QC Bible

\bibitem{Nielsen} M. A. Nielsen and I. L. Chung, Quantum Computation and Quantum Information, Cambridge University Press (2010). 

%Shor

\bibitem{Shor} ``Polynomial-Time Algorithms for Prime Factorization and Discrete Logarithms on a Quantum Computer", P. W. Shor, SIAM J. Comput., 26 (5): 1484-1509, (1997)

\bibitem{Orus_2004} ``Universality of entanglement and quantum-computation complexity'', R. Orus and J. I. Latorre, Phys. Rev. A {\bf 69}, 052308 (2004)

% ME 

\bibitem{vanMeter_2005} ``Fast Quantum Modular Exponentiation'', R. Van Meter, K. M. Itoh, Phys. Rev. A {\bf 71}, 052320 (2005)

\bibitem{Pavlidis_2013} ``Fast Quantum Modular Exponentiation Architecture for Shor's Factorization Algorithm'', A. Pavlidis, D. Gizopoulos, Quantum Information and Computation, Vol. {\bf 14}, No. 7\&8 0649-0682 (2014) 


% holographic tree properties

\bibitem{Orus_2014} ``Advances on Tensor Network Theory: Symmetries, Fermions, Entanglement, and Holography", R. Orus, Eur. Phys. J. B 87: 280 (2014)

\bibitem{Gottesman_97} D. Gottesman, Caltech Thesis, arXiv:quant-ph/9705052, (1997)



\end{thebibliography}
\end{document}